\documentclass[twocolumn,pra,superscriptaddress]{revtex4}
\usepackage{amsmath}
\usepackage{amssymb}
\usepackage{graphicx}
\usepackage{color}
\usepackage{hyperref}
\usepackage{bbm}
\usepackage{enumerate}

\begin{document}

\title{Pseudo-Hermitian Hamiltonians generating waveguide mode evolution}

\author{Penghua Chen}

\affiliation{Division of Physics and Applied Physics, School of
  Physical and Mathematical Sciences, Nanyang Technological
  University, Singapore 637371, Singapore}

\author{Y.~D.~Chong}
\email{yidong@ntu.edu.sg}

\affiliation{Division of Physics and Applied Physics, School of
  Physical and Mathematical Sciences, Nanyang Technological
  University, Singapore 637371, Singapore}

\affiliation{Centre for Disruptive Photonic Technologies, Nanyang
  Technological University, Singapore 637371, Singapore}

\date{\today}

\begin{abstract}
We study the properties of Hamiltonians defined as the generators of
transfer matrices in quasi-one-dimensional waveguides.  For single- or
multi-mode waveguides obeying flux conservation and time-reversal
invariance, the Hamiltonians defined in this way are non-Hermitian,
but satisfy symmetry properties that have previously been identified
in the literature as ``pseudo Hermiticity'' and ``anti-PT
symmetry''.  We show how simple one-channel and two-channel models
exhibit transitions between real, imaginary, and
complex eigenvalue pairs.
\end{abstract}

\maketitle

\section{Introduction}
\label{sec:Introduction}

In 1998, Bender and co-workers \cite{Bender,Bender2,Berry} pointed out
that Hamiltonians which are symmetric under a combination of parity
and time reversal (PT) possess an interesting feature: despite being
non-Hermitian, they can have eigenvalues that are strictly real.
Moreover, tuning the Hamiltonian parameters can induce a non-Hermitian
symmetry-breaking transition, between PT-unbroken eigenstates with
real eigenvalues and PT-broken eigenstates with complex and
conjugate-paired eigenvalues.  The original intention of Bender
\textit{et al.}~was to use PT symmetry to extend fundamental quantum
mechanics, but in 2008 Christodoulides and co-workers showed that PT
symmetry could be realized in optical structures with balanced gain
and loss \cite{Makris,Musslimani,Guo,Ruter}.  Since then, research
into PT symmetric optics has progressed rapidly, and the idea of
``gain and loss engineering'' in photonics, based on PT symmetry, has
led to devices with highly promising applications, such as low-power
optical isolation \cite{Peng,Chang} and laser mode selection
\cite{PTLaser1,PTLaser2}.

In this paper, we look at a class of physically-motivated
non-Hermitian Hamiltonians that are \textit{not} PT symmetric in the
original sense of Bender \textit{et al.}~\cite{Bender,Bender2,Berry},
but nonetheless exhibit symmetry-breaking transitions between real and
complex eigenvalues.  These Hamiltonians are the generators of
transfer matrices in single- or multi-channel waveguides
\textit{without} gain or loss, at a fixed frequency (or energy).  They
were first studied in a 1997 paper by Mathur \cite{mathur1997}, prior
to the development of PT symmetry; one of our goals is to re-evaluate
them in light of subsequent developments in the theory of
non-Hermitian Hamiltonians.  We show that they form a subset of the
``pseudo-Hermitian'' matrices, a class of non-Hermitian matrices
identified by Mostafazadeh and co-workers as a generalization of the
PT symmetry concept
\cite{mostafa2002,mostafa2003,mostafa2004,mostafa2010,mostafa2014-1,
  mostafa2014-2, mostafa2016}.  This subset is restricted further by a
generalized anti-PT symmetry
\cite{PengPeng2016,Sukhorukov2015,LiGe2013,Wu2015}, where the $P$
operation interchanges forward-going and backward-going waveguide
modes.  A similar anti-PT symmetry was previously identified by
Sukhorukov \textit{et al.}~in the context of beam evolution in
parametric amplifiers \cite{Sukhorukov2015}.  Mostafazadeh and
co-workers have also studied the Hamiltonians that generate transfer
matrices in waveguides containing gain and/or loss, i.e.~without flux
conservation and time-reversal invariance \cite{mostafa2014-1,
  mostafa2014-2, mostafa2016}.  When those symmetries \textit{are}
present, however, we show that the resulting pseudo-Hermiticity and
anti-PT symmetries give rise to non-Hermitian transitions between
purely imaginary, purely real, and complex values.  These
non-Hermitian transitions are reminiscent of PT symmetry breaking
transitions, and we show that they have physical consequences for the
transmission properties of single- and multi-mode waveguides under
parameter change.

For an $N$-channel waveguide (either an optical waveguide, or a
quantum electronic waveguide \cite{mathur1997}) that obeys flux
conservation as well as time-reversal invariance, it is known
\cite{MPK} that the group of transfer matrices, at a given operating
frequency or energy, has a one-to-one mapping to the symplectic group
$\mathrm{Sp}(2N,\mathbb{R})$.  The transfer matrices are generated by
$2N\times 2N$ matrices that are typically non-Hermitian; we can regard
each such generator, $H$, as a Hamiltonian.  The $H$ matrices map onto
the group of real matrices of the ``Hamiltonian'' type,
$\mathrm{sp}(2N,\mathbb{R})$, which are the generators of
$\mathrm{Sp}(2N,\mathbb{R})$.

The eigenvalues of $H$ are not energies or frequencies, but rather the
modal wavenumbers of a translationally invariant waveguide.  Real
eigenvalues correspond to propagating modes, and complex eigenvalues
correspond to evanescent (in-gap) modes.  As shown below, $H$ supports
real eigenvalues despite being non-Hermitian because it satisfies a
certain pair of symmetries: pseudo-Hermiticity
\cite{mostafa2002,mostafa2003,mostafa2004,mostafa2010} and anti-PT
symmetry \cite{PengPeng2016,Sukhorukov2015,LiGe2013,Wu2015}.  
These symmetries are tied to the physical conditions of flux conservation
and time-reversal invariance in the underlying waveguide.

To motivate the interpretation of transfer matrix generators as
``Hamiltonians'' \cite{mathur1997, mostafa2014-1, mostafa2014-2,
  mostafa2016}, consider a segment of an $N$-channel waveguide with
negligible back-reflection.  The position along the waveguide axis,
$z$, can be thought of as playing the role of ``time''.  At a given
energy $E$, the transfer matrix is a $2N\times2N$ block-diagonal
matrix of the form
\begin{equation}
  M = \begin{pmatrix} U_1 & 0 \\ 0 & U_2 \end{pmatrix},
\end{equation}
where $U_1$ and $U_2$ describe the mode-mixing of the forward- and
backward-going modes, respectively.  In the absence of gain or loss,
$U_1$ and $U_2$ are unitary, and $M$ is generated by a $2N\times2N$ Hermitian matrix $H$ such that
\begin{equation}
  M = \exp(iHz) = \begin{pmatrix} e^{iH_1z} & 0 \\ 0 & e^{iH_2z} \end{pmatrix},
\end{equation}
where $H_1$ and $H_2$ are Hermitian sub-matrices of $H$. The eigenvalues of $H_{1,2}$ are the wavenumbers of the forward- and
backward-going modes.  For such a reflection-free waveguide, we could
focus on the Hermitian submatrix $H_1$ as the Hamiltonian of the $N$
forward modes; this leads to the well-known mapping between beam
propagation and the Schr\"odinger wave equation \cite{DeRaedt}.

When back-reflection is non-negligible, due to inhomogeneities in the
waveguide (e.g., a fiber Bragg grating
\cite{Mizrahi1993,Erdogan1996,Sukhorukov2007}), $M$ is no longer
block-diagonal; its generator $H$ is neither block-diagonal nor
Hermitian \cite{mathur1997}.  However, the eigenvalues of $H$
can still be regarded as modal wavenumbers, which now consist of a mix
of forward-going and backward-going components.  Band extrema
correspond to exceptional points of $H$, where its eigenvectors
coalesce and the matrix becomes defective.  At these points, the modal
wavenumbers exhibit ``symmetry-breaking'' transitions between real
pairs (propagating modes), and either purely imaginary pairs (purely
evanescent gap modes) or complex pairs (quasi-evanescent gap modes).
This is reminiscent of PT symmetry breaking \cite{Bender,Bender2},
but, as we shall see, it is not PT symmetry that is responsible for
these eigenvalue transitions.

\section{Hamiltonian Symmetries}
\label{sec:H_symmetry}

We begin with a brief summary of the definitions of PT symmetry and
some of its variants and generalizations
\cite{Wang2012,Wang2010,Bender3}.  First, a Hamiltonian $H$ is ``PT
symmetric'' if it is invariant under a combination of a unitary parity
operator $P$ and an antiunitary time-reversal operator $T$ (which we
take to be the complex conjugation operator), as follows
\cite{Bender,Bender2,Berry}:
\begin{equation}
  H=(PT)H(PT)^{-1}.
  \label{PTsym}
\end{equation} 
Conventionally, a parity operator must be involutory ($P^2 = I$), but
we can generalize the PT symmetry concept by dropping this assumption,
and requiring only the combined antiunitary operator $PT$ to be
involutory \cite{Berry,Wang2012,Wang2010,Bender3}.  Eq.~(\ref{PTsym}) then
implies that eigenvalues of $H$ are either purely real, or form
complex conjugate pairs.  Next, ``pseudo-Hermiticity'' is a slightly
different concept from PT symmetry
\cite{mostafa2002,mostafa2003,mostafa2004,mostafa2010}: given a linear
invertible Hermitian operator $\eta$, a Hamiltonian $H$ is said to be
pseudo-Hermitian under $\eta$ if
\begin{equation}
  H^{\dagger}=\eta H\eta^{-1}.
\end{equation}
The $\eta$ operator then serves as the metric operator for a possibly
indefinite inner product.  If $H$ is a PT symmetric matrix whose
eigenvalues are all real, then $H$ is necessarily pseudo-Hermitian
under some operator $\eta$, but not vice versa
\cite{mostafa2002,mostafa2003,mostafa2004,mostafa2010}.  Finally, we
say that Hamiltonian is ``anti-PT symmetric'' if
\begin{equation}
  -H=PTH(PT)^{-1}.
\end{equation}
In this case, the eigenvalues are either purely imaginary, or occur in
negative-conjugate pairs
\cite{LiGe2013,Sukhorukov2015,Wu2015,PengPeng2016}.  A physical
example of such a symmetry can be found in parametric amplifiers,
where the $P$ operator interchanges signal and idler waveguide
channels \cite{Sukhorukov2015}.

We now consider a waveguide that supports $N$ channels (waveguide
modes), operating at a single fixed frequency or energy.  At each
position $z$ along the waveguide axis, the wavefunction can be
expressed by $2N$ complex wave amplitudes:
\begin{equation}
  |\Psi(z)\rangle \equiv \begin{pmatrix} \Psi^+(z) \\ \Psi^-(z)
  \end{pmatrix},\;\; \mathrm{where}\;\;
  \Psi^\pm(z) \equiv \begin{pmatrix}
    \psi_{1}^\pm\left(z\right) \\ \vdots \\\psi_{N}^\pm(z)
  \end{pmatrix}.
\end{equation}
Here, $\pm$ denotes wave components moving in the $\pm\hat{z}$
direction, and the wave components are normalized so that
$|\psi_n^\pm|^2$ is an energy flux.  The wavefunctions at any two
points, $z_1$ and $z_2$, are related by a transfer matrix:
\begin{equation}
  M(z_1,z_2)\, |\Psi(z_2)\rangle = |\Psi(z_1)\rangle.
\end{equation}
Let us assume that the waveguide is flux-conserving and time-reversal
invariant \cite{MPK}.  This imposes two symmetry constraints on $M$.
First, flux conservation states that the incoming flux into the
segment between $z_1$ and $z_2$ must equal the outgoing flux, which
implies that
\begin{equation}
  \Sigma_z = M^\dagger \Sigma_z M, \;\;\mathrm{where}\;\; \Sigma_z
  \equiv \begin{pmatrix} \mathcal{I} & \mathbf{0} \\ \mathbf{0}
    & -\mathcal{I}
  \end{pmatrix}.
  \label{Msymmetry1}
\end{equation}
Secondly, time-reversal invariance states that for each solution,
there is also a solution obtained by taking the complex conjugate of
the wavefunctions.  Hence,
\begin{equation}
  M^* = \Sigma_x M \Sigma_x, \;\;\mathrm{where}\;\; \Sigma_x
  \equiv \begin{pmatrix} \mathbf{0} & \mathcal{I} \\ \mathcal{I} &
    \mathbf{0}
  \end{pmatrix},
  \label{Msymmetry2}
\end{equation}
with $\mathcal{I}$ denoting the $N\times N$ identity matrix.  Both of
these symmetry relations are preserved under composition of transfer
matrices.
===
The combination of Eqs.~(\ref{Msymmetry1}) and (\ref{Msymmetry2})
implies that waveguide propagation is reciprocal.  Note that
Eq.~(\ref{Msymmetry1}) can be re-written as $M\Sigma_z M^\dagger =
\Sigma_z$.  We can combine this with Eq.~(\ref{Msymmetry2}) to obtain
\begin{equation}
  M^TJ M = J,
  \label{MJM}
\end{equation}
where
\begin{equation}
  J \equiv \Sigma_z \Sigma_x =
  \begin{pmatrix} \mathbf{0} &\mathcal{I} \\ -\mathcal{I} & \mathbf{0}
  \end{pmatrix}.
  \label{jmatrix}
\end{equation}
This implies that the waveguide is reciprocal \cite{reciprocity}; to
see this, consider two arbitrary independent sets of wave amplitudes
$\psi_A^\pm$ and $\psi_B^\pm$, such that
\begin{equation}
  M \begin{pmatrix}\Psi_A^+ \\ \Psi_A^- \end{pmatrix}
  = \begin{pmatrix}\Phi_A^+ \\ \Phi_A^- \end{pmatrix}, \;\;
  M \begin{pmatrix}\Psi_B^+ \\ \Psi_B^- \end{pmatrix}
  = \begin{pmatrix}\Phi_B^+ \\ \Phi_B^- \end{pmatrix}.
  \label{arbitrary_amplitudes}
\end{equation}
We can also define the scattering matrix $S$, which relates incoming
to outgoing waves:
\begin{equation}
  S \begin{pmatrix}\Psi^+ \\ \Phi^- \end{pmatrix}
  = \begin{pmatrix}\Psi^- \\ \Phi^+ \end{pmatrix},
  \label{Smatrix}
\end{equation}
for both $A$ and $B$ subscripts.
By applying Eqs.~(\ref{MJM})--(\ref{arbitrary_amplitudes}) to
Eq.~(\ref{Smatrix}), we can show that
\begin{equation}
  \begin{pmatrix}\Psi_A^+ \\ \Phi_A^- \end{pmatrix}^T (S - S^T)
  \begin{pmatrix}\Psi_B^+ \\ \Phi_B^- \end{pmatrix} = 0.
\end{equation}
Since this holds for independent sets of wave amplitudes, we conclude
that $S$ must be symmetric \cite{reciprocity}.  It is important to
note, however, that Eqs.~(\ref{Msymmetry1})--(\ref{Msymmetry2})
together form a set of constraints that is stronger than just the
reciprocity condition (\ref{MJM}).  For instance, in optical
waveguides with gain and/or loss,
Eqs.~(\ref{Msymmetry1})--(\ref{Msymmetry2}) are violated, but
Eq.~(\ref{MJM}) still holds.

We can now define a ``Hamiltonian'' $H$ that is the infinitesimal
generator of the transfer matrix, via the Schr\"odinger-like equation
\begin{align}
  \begin{aligned}
    -i\frac{\partial}{\partial z}M(z,z_{0}) &= H(z)\; M(z,z_{0}) \\
    \Leftrightarrow \;\;\; -i\frac{\partial}{\partial z}|\Psi(z)\rangle &=
    H(z) |\Psi(z)\rangle.
  \end{aligned}
  \label{eq:Mathur's equation}
\end{align}
The goal of this paper is to understand and interpret the symmetry
properties of $H$.  This matrix is Hermitian if and only if $M$ is
unitary, which corresponds to the case of the reflection-free
waveguide discussed in Section~\ref{sec:Introduction}.

In the more general case where $M$ is non-unitary, $H$ is
non-Hermitian.  To determine the symmetry constraints on $H$, we use
the well-known fact that the exponential map commutes with the adjoint
action:
\begin{equation}
  e^{J^{-1}\left(iH\ensuremath{\Delta}x\right)J} = J^{-1}e^{iH\ensuremath{\Delta}x}J,
\label{eq:Exponential map commutes with adjoint action}
\end{equation}
where $J$ is defined in Eq.~(\ref{jmatrix}). Applying this to Eqs.~(\ref{Msymmetry1})--(\ref{Msymmetry2}) gives the
following pair of symmetry relations for $H$:
\begin{align}
  \varSigma_{z}H\varSigma_{z} &= H^{\dagger}.
  \label{eq:hamiltonian_flux} \\
  \varSigma_{x}H\varSigma_{x} &= -H^{*}.
  \label{eq:hamiltonian_T}
\end{align}
Based on the definitions introduced at the beginning of this section,
$H$ is pseudo-Hermitian under the operator $\varSigma_{z}$
\cite{mostafa2014-2}, and anti-PT symmetric under the operator
$\varSigma_x$.

A $2N\times2N$ matrix $H$ satisfies these two symmetries,
(\ref{eq:hamiltonian_flux}) and (\ref{eq:hamiltonian_T}), if and only
if it has the form
\begin{equation}
  H = \begin{pmatrix}
    \mathcal{H} & \mathcal{A}\\
    -\mathcal{A}^{*} & -\mathcal{H}^{*}
  \end{pmatrix},
  \label{eq:Hamiltonian form}
\end{equation}
where $\mathcal{H}$ and $\mathcal{A}$ are $N\times N$ matrices satisfying
\begin{equation}
  \mathcal{H} = \mathcal{H}^\dagger, \quad \mathcal{A} = \mathcal{A}^T.
  \label{eq:Hamiltonian form 2}
\end{equation}

These $H$ matrices are closely connected to the symplectic structure
of the transfer matrices.  It is known that the transfer matrices can
be mapped to the real symplectic group $\mathrm{Sp}(2N,\mathbb{R})$
\cite{MPK}.  In a similar way, we can show that the $H$ matrices are
isomorphic (in the vector space sense) to the real-valued Hamiltonian
matrices, $\mathrm{sp}(2N,\mathbb{R})$, which are the Lie algebra
generators of $\mathrm{Sp}(2N,\mathbb{R})$.  To prove this, we first
note, via Eqs.~(\ref{Msymmetry1})--(\ref{Msymmetry2}), that the
transfer matrices take the form
\begin{equation}
  M = \begin{pmatrix}
    \mathcal{C} & \mathcal{B}\\
    \mathcal{B}^{*} & \mathcal{C}^{*}
  \end{pmatrix},
  \label{eq:Meq1}
\end{equation}
where $\mathcal{B}$ and $\mathcal{C}$ are complex $N\times N$ matrices
satisfying
\begin{equation}
  \mathcal{C}\mathcal{C}^{\dagger}-\mathcal{B}\mathcal{B}^{\dagger}=1 \;\;
  \text{and} \;\;
  \mathcal{C}\mathcal{B}^{T}=\mathcal{B}^{T}\mathcal{C}.
  \label{eq:Meq2}
\end{equation}
Define $\mathcal{C}=\mathcal{X}+i\mathcal{Y}$ and
$\mathcal{B}=\mathcal{F}+i\mathcal{G}$, where
$\{\mathcal{X},\mathcal{Y},\mathcal{F},\mathcal{G}\}$ are real
$N\times N$ matrices.  Then $M$ maps to a real $2N\times2N$ matrix as
follows \cite{MPK}:
\begin{equation}
  f(M)=W = \begin{pmatrix} \mathcal{X}-\mathcal{G} & \mathcal{F}-\mathcal{Y}
    \\ \mathcal{F}+\mathcal{Y} & \mathcal{X}+\mathcal{G} \end{pmatrix}.
  \label{fmap}
\end{equation}
The $f$ map is one-to-one and onto, and one can show that $W$ is
symplectic (i.e., $W \, J W^{T} = J$) if and only if $M$ satisfies
Eqs.~(\ref{eq:Meq1})--(\ref{eq:Meq2}). Note, however, that the group
operation of $\mathrm{Sp}(2N,\mathbb{R})$---i.e., multiplication of
the $W$ matrices---does not correspond to the composition operation
(matrix multiplication) of the transfer matrices.

The map $f$ defined in Eq.~(\ref{fmap}) can also be applied to the $H$
matrices, which are the generators of $M$ satisfying
Eqs.~(\ref{eq:Hamiltonian form})--(\ref{eq:Hamiltonian form 2}).  We
can then show that
\begin{equation}
if(H) =
\begin{pmatrix} 
-\mathrm{Im}(\mathcal{H})-\mathrm{Re}(\mathcal{A})
& -\mathrm{Re}(\mathcal{H})-\mathrm{Im}(\mathcal{A})\\
\mathrm{Re}(\mathcal{H})-\mathrm{Im}(\mathcal{A})
& -\mathrm{Im}(\mathcal{H})+\mathrm{Re}(\mathcal{A})
\end{pmatrix},
\label{Hmapping}
\end{equation}
which is a real $2N\times2N$ matrix of the ``Hamiltonian'' form.  This
means that
\begin{align}
  [Jf(H)]^{T}=Jf(H),
\end{align}
where $J$ is the skew-symmetric matrix defined in Eq.~(\ref{jmatrix}).
(The Hamiltonian matrices are so-called because they occur naturally
in systems of equations formed by Hamilton's equations of classical
mechanics \cite{meyer2008}; the terminology is unrelated
to our interpretation of the $H$ matrices as ``Hamiltonians'', which
is based on the Schr\"odinger-like Eq.~(\ref{eq:Mathur's equation}).)
Eq.~(\ref{Hmapping}) is one-to-one, and thus constitutes an
isomorphism to the group of Hamiltonian matrices, where the group
operations are addition operations on the $H$ matrices as well as the
Hamiltonian matrices. In turn, the group of Hamiltonian matrices is
the Lie algebra $\mathrm{sp}(2N, \mathbb{R})$ that generates the Lie
group $\mathrm{Sp}(2N, \mathbb{R})$.

\section{Eigenvalue properties}
\label{Eigenvalue properties}

The $H$ matrices defined in the previous section are pseudo-Hermitian
under $\varSigma_{z}$, due to the flux conservation condition
(\ref{eq:hamiltonian_flux}), and anti-PT symmetric under
$\varSigma_x$, due to the time-reversal invariance condition
(\ref{eq:hamiltonian_T}).  These symmetries offer the prospect of
non-Hermitian transitions in the eigenvectors and eigenvalues of $H$,
analogous to the PT-breaking transition.

First, however, let us discuss the physical meaning of the
eigenvectors and eigenvalues of $H$.  If a waveguide is invariant
under translation by some $L$, each waveguide mode is describable by a
state vector $|\Psi\rangle$ satisfying \cite{floquet}
\begin{equation}
  M(z+L,z)\, |\Psi\rangle = \exp(i\kappa L) |\Psi\rangle,
  \label{Mequation}
\end{equation}
for some wavenumber $\kappa$.  Then $|\Psi\rangle$ and $\kappa$ are an
eigenvector and eigenvalue of $H$, the generator of $M$.  The
components of the state vector $|\Psi\rangle$ describe the waveguide
mode's decomposition into forward-going and backward-going modes of an
``empty'' (reflection-free) waveguide.  These forward and backward
modes are coupled via inhomogeneities in the waveguide, such as a
Bragg grating where $L$ is a multiple of the grating period
\cite{Mizrahi1993,Erdogan1996,Sukhorukov2007}.

The Hamiltonian $H$ satisfies
Eqs.~(\ref{eq:hamiltonian_flux})--(\ref{eq:hamiltonian_T}), which
implies that its eigenvalues come in pairs.  If $\kappa$ is an
eigenvalue, then $\kappa^{*}$ and $-\kappa^*$ are also eigenvalues.
To show this, suppose that
\begin{equation}
  H\begin{pmatrix}v \\ w \end{pmatrix} = \kappa \begin{pmatrix}v \\ w \end{pmatrix}
\end{equation}
for some $v, w \in \mathbb{C}^N$ and $\kappa \in \mathbb{C}$.  By the
pseudo-Hermiticity relation (\ref{eq:hamiltonian_flux}),
\begin{equation}
  H^\dagger \begin{pmatrix}v \\ -w \end{pmatrix} = \kappa \begin{pmatrix}v \\ -w \end{pmatrix}.
  \label{eq:Heigenval1}
\end{equation}
Since $H$ and $H^{T}$ share the same eigenvalues, $\kappa^*$ is an
eigenvalue of $H$.  Furthermore, from the anti-PT symmetry relation
(\ref{eq:hamiltonian_T}), we obtain
\begin{equation}
  H^* \begin{pmatrix}w \\ v \end{pmatrix} = -\kappa \begin{pmatrix}w \\ v \end{pmatrix},
  \label{eq:Heigenval2}
\end{equation}
which implies that $-\kappa^*$ is an eigenvalue of $H$.  Based on
these results, for each eigenvalue $\kappa$, either
\begin{enumerate}[(i)]
\item $\kappa$ is purely imaginary;
\item $\kappa$ is purely real; or
\item there exist four eigenvalues $\{\kappa, -\kappa, \kappa^*,
  -\kappa^*\}$ that are not purely real or imaginary.
\end{enumerate}
Note that case (iii) can only occur for $N \ge 2$.

We can induce transitions between these three cases by tuning various
parameters in $H$.  The transition
from case (i) to case (ii) or (iii) corresponds to the breaking of the
eigenvector's anti-PT symmetry; in case (i), the eigenvector has
unbroken anti-PT symmetry and satisfies $|v|^2 = |w|^2$ [see
  Eq.~(\ref{eq:Heigenval2})], whereas in the other two cases, the
eigenvector's anti-PT symmetry is broken.  However, the transition
from (ii) to (iii) does \textit{not} seem to correspond to any obvious
form of symmetry-breaking (or ``pseudo-Hermiticity breaking'') in the
eigenvector, based on Eq.~(\ref{eq:Heigenval1}).

\section{Implications for waveguide scattering parameters}

In PT symmetric optics, the evolution of waveguide modes can be used
to provide evidence for the PT symmetry-breaking transition
\cite{Makris,Musslimani,Guo,Ruter}.  In the pioneering experimental
demonstration of Ruter \textit{et al.}, for instance, light is
injected into one of a pair of PT symmetric waveguides (which are
assumed to have negligible back-reflection) \cite{Ruter}.  When the
system is in the PT-unbroken phase, the injected light beats, or
oscillates, between the two waveguides with no net amplification or
damping; but when the system is in the PT-broken phase, the injected
light experiences exponential amplification.

In a similar spirit, we will now show that the non-Hermitian
transitions of $H$ are tied to the modal evolution across a waveguide
segment with \textit{non}-negligible back-reflection.  This process is
interpreted through the \textit{backward} $z$-evolution of the wave
amplitudes.  Unlike in Ref.~\onlinecite{Ruter}, the different
components of the state vector in our model consist of forward-going
versus backward-going waves, rather than the excitations of gain
versus loss waveguides.  We can make use of the fact that, at the end
of the waveguide, the transmitted wave is purely forward-going, with
no backward component; this plays a role analogous to the initial
state in the experiment of Ref.~\onlinecite{Ruter}, in which light is
injected into one of the two waveguides.  Our non-Hermitian
transitions correspond to the abrupt changes in reflection and
transmission caused by passing through band extrema.

For a single-mode waveguide ($N=1$), the $H$ matrix has the form
\begin{equation}
  H=\begin{pmatrix} \mathcal{E} & a\\ -a^{*} & -\mathcal{E} \end{pmatrix},
  \label{H22}
\end{equation}
where $\mathcal{E} \in \mathbb{R}$ and $a \in\mathbb{C}$.  In the
context of an optical waveguide, $\mathcal{E}$ and $a$ could
parameterize the detuning and back-reflection induced by a fiber Bragg
grating \cite{PengPeng2016,Sukhorukov2015,LiGe2013,Wu2015}.
Interestingly, a non-Hermitian Hamiltonian of this form has previously
been investigated by Mathur \cite{mathur1997}, in the context
of a quantum electronic waveguide formed by coupled chiral edge states
in a quasi-one-dimensional quantum Hall gas.  There, the chiral edge
states satisfy the time-independent Schr\"odinger equation
\begin{equation}
  \begin{pmatrix} -i\partial_z & -a(z) \\ -a(z)^* & i\partial_z
  \end{pmatrix}
  \begin{pmatrix}\psi^+\\ \psi^-
  \end{pmatrix}
  = \mathcal{E}
  \begin{pmatrix}\psi^+\\ \psi^-
  \end{pmatrix},
  \label{mathur_model}
\end{equation}
where $\psi^\pm$ are the wave amplitudes for the chiral edge states on
opposite edges, $\mathcal{E}$ is the edge state energy, and $a(z)$ is
the coupling between the edge states.  (If the sample is sufficiently
narrow relative to the penetration depth of the edge states, such
coupling can occur by evanescent tunneling; it spoils the topological
protection that is normally enjoyed by the edge states
\cite{Hatsugai,Hailong}).  Re-arranging Eq.~(\ref{mathur_model}), and
using the definition (\ref{eq:Mathur's equation}), yields
Eq.~(\ref{H22}).

When $a$ is a non-zero constant, $H$ has eigenvalues
\begin{equation}
  \kappa=\pm\sqrt{\mathcal{E}^2- |a|^2}.
\end{equation}
The eigenvalues are either both imaginary (corresponding to unbroken
anti-PT symmetry), or both real (corresponding to broken anti-PT
symmetry).  The transition between the two regimes occurs at $\kappa =
0$, and it corresponds to the familiar effects of crossing a band
extremum, or of crossing the cutoff frequency of a waveguide.  On one
side of the transition, there is a pair of propagating modes (real
$\kappa$), and on the other side the modes are purely evanescent
(imaginary $\kappa$).  Since there are only two eigenvalues, this
anti-PT-breaking transition is the only type that occurs for $N=1$.

\begin{figure}
\centering
\includegraphics[width=1.0\columnwidth]{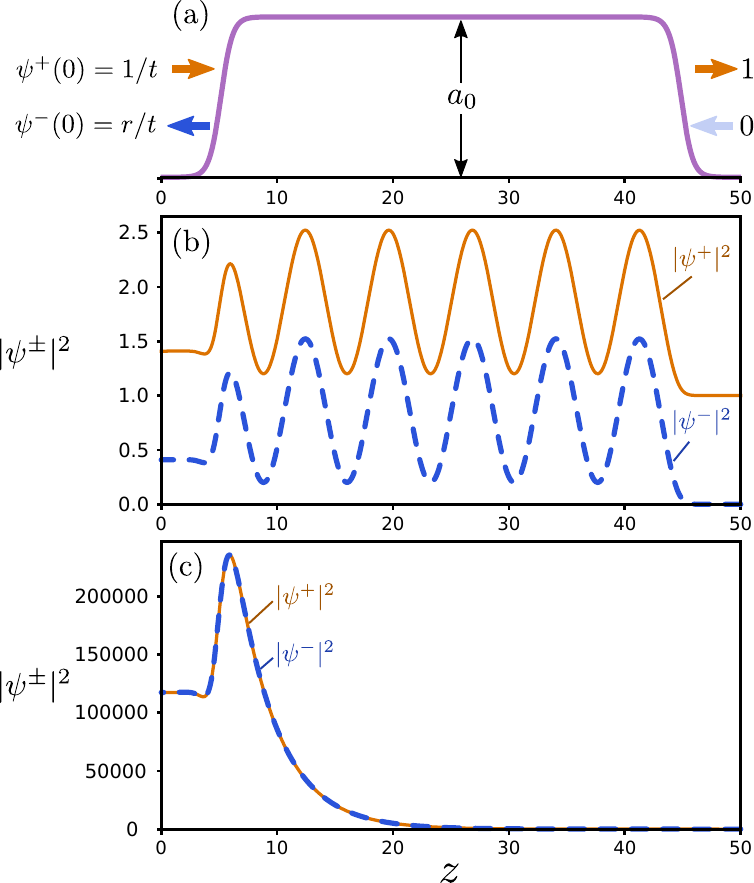}
\caption{Reflection and transmission for a waveguide segment.  (a)
  Schematic of the scattering problem, showing input wave amplitude
  $1/t$, reflected amplitude $r/t$, and transmitted amplitude $1$.
  The curve shows the profile of $a(z)$, which is taken to be $a(z) =
  (a_0/2)[\tanh(z-z_1) - \tanh(z-z_2)]$, where $z_1 = 5$ and $z_2 =
  45$.  We set $\mathcal{E} = 1$ everywhere, and $a = 0$ outside the
  segment.  (b)--(c) Intensities of the forward and backward
  components, $|\psi^\pm(z)|^2$, versus $z$, for (b) $a_0 = 0.9$
  (anti-PT-broken), and (c) $a_0 = 1.01$ (anti-PT-unbroken).}
\label{fig:n1plots}
\end{figure}

In Fig.~\ref{fig:n1plots}, we consider a waveguide segment between $z
= 0$ and $z = Z$.  Within the segment, the parameter $a$ varies
smoothly from zero (at the end-points) to a non-zero value $a_0$, as
shown in Fig.~\ref{fig:n1plots}(a), whereas outside the segment, we
set $a = 0$ corresponding to a reflection-free waveguide; we also set
$\mathcal{E} = 1$ everywhere.  Suppose that there is an input wave
with amplitude $\psi^+(0) = 1$ at the start of the segment.  This
gives rise to a reflected wave with amplitude $\psi^-(0) = r$, and a
transmitted wave with amplitude $\psi^+(Z) = t$.  Since there is no
backward-going wave at the end of the segment, $\psi^-(Z) = 0$.  Due
to linearity, we can rescale the wave amplitudes so that $\psi^+(0) =
1/t, \psi^-(0) = r/t$ at the start of the segment, and $\psi^+(Z) = 1,
\, \psi^-(Z) = 0$ at the end.

This reflection-and-transmission scenario can be mathematically
described by the \textit{backward} evolution of the non-Hermitian
Schr\"odinger-like equation (\ref{eq:Mathur's equation}), with the end
conditions $\psi^+(Z) = 1$ and $\psi^-(Z) = 0$.  Upon integrating
Eq.~(\ref{eq:Mathur's equation}) backwards from $z = Z$ to $z = 0$, we
obtain $\psi^+(0) = 1/t$ and $\psi^-(0) = r/t$.  In
Figs.~\ref{fig:n1plots}(b)--(c), we plot $|\psi^+(z)|^2$ and
$|\psi^-(z)|^2$ against $z$, obtained through the backward evolution
calculation described above.  When $a_0 = 0.9$, the entire waveguide
segment is in the anti-PT-broken phase, where the eigenvalues of $H$
are real; we observe beating between the forward-going and
backward-going components, without exponential amplification or
damping.  When $a_0 = 1.01$, the waveguide segment passes into the
anti-PT-unbroken phase, where the eigenvalues of $H$ are imaginary; we
observe that the state undergoes exponential ``backward
amplification'' in going from $z = Z$ to $z = 0$.  Interpreted in
terms of the forward-going wave injected at $z=0$, the transmitted
part is exponentially damped because, inside the segment, the
waveguide passes through a band extremum (into a band-gap, or under
the waveguide cutoff).  Most of the incident wave is hence reflected.
The two distinct behaviors---oscillation and damping---correspond to
the anti-PT-broken and anti-PT-unbroken phases of the non-Hermitian
Hamiltonian $H$.  This is in some sense the opposite of the PT
waveguide evolution experiment of Ruter \textit{et al.}, where the
symmetry-unbroken regime produces oscillations and the symmetry-broken
regime produces amplification~\cite{Ruter}.

\begin{figure}
\centering
\includegraphics[width=1.0\columnwidth]{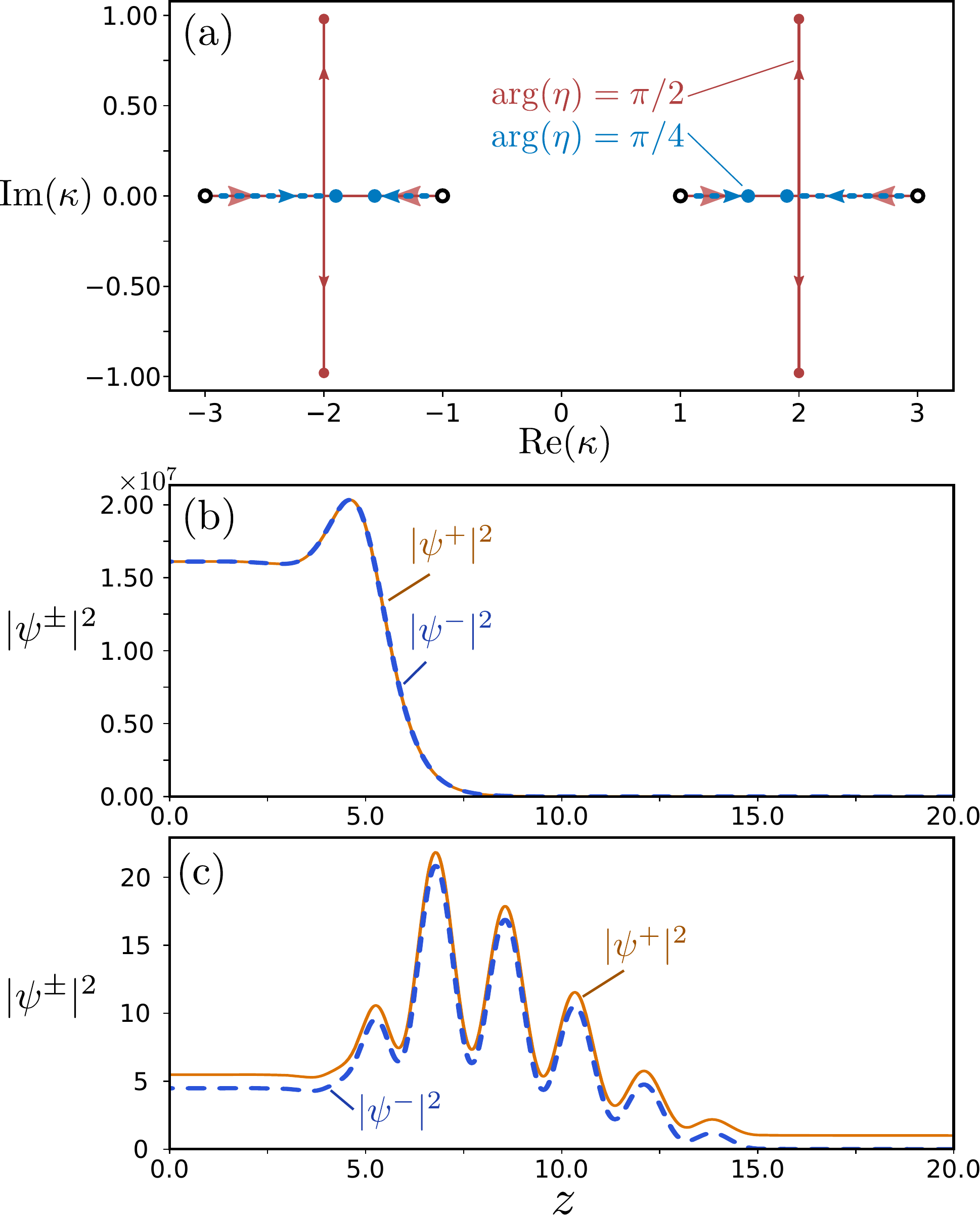}
\caption{Reflection and transmission behavior of an $N=2$ waveguide.
  (a) Complex plane trajectories of the eigenvalues $\kappa$, for the
  matrix given in Eq.~(\ref{H22a}) with $\mathcal{E} = 1$, $|\eta| =
  2$, $a = 0$, and $m_1 = m_2 = m$ varying from 0 to 1.4.  The
  trajectory directions are indicated by arrows, the start points ($m
  = 0$) are indicated by hollow circles, and the end points ($m =
  1.5$) are indicated by filled circles.  The trajectories are
  dependent on $\mathrm{arg}(\eta)$, and two possible choices, $\pi/4$
  and $\pi/2$, are shown.  (b)--(c) Intensities of the forward and
  backward components, $|\psi^\pm(z)|^2 = |\psi_\uparrow^\pm(z)|^2 +
  |\psi_\downarrow^\pm(z)|^2$, versus $z$.  We set $\mathcal{E} = 1$,
  $|\eta| = 2$, $a = 0$, and $m = (1.4/2)[\tanh(z-z_1) -
    \tanh(z-z_2)]$, where $z_1 = 5$ and $z_2 = 15$.  (b) For
  $\mathrm{arg}(\eta) = \pi/4$, the eigenvalues become complex within
  the waveguide segment, resulting in exponential wave damping.  (c)
  For $\mathrm{arg}(\eta) = \pi/2$, the eigenvalues are real
  throughout, resulting in beating.  The end-point condition is
  $\psi_\uparrow^+(20) = 1$; varying the distribution of forward-going
  wave amplitudes at the end-point does not significantly alter the
  results.  }
\label{fig:n2plots}
\end{figure}

Next, we consider the $N = 2$ case.  The most general $4\times 4$
non-Hermitian Hamiltonian satisfying Eqs.~(\ref{eq:Hamiltonian
  form})--(\ref{eq:Hamiltonian form 2}) can be written as
\begin{equation}
  H = \begin{pmatrix}
    \mathcal{E} & \eta & m_1 & a\\
    \eta^{*} & \mathcal{E} & a & m_2\\
    -m_1^{*} & -a^{*} & -\mathcal{E} & -\eta^{*}\\
    -a^{*} & -m_2^{*} & -\eta & -\mathcal{E}
  \end{pmatrix},
  \label{H22a}
\end{equation}
where $\mathcal{E} \in \mathbb{R}$ and $\eta, a, m_1, m_2 \in
\mathbb{C}$.  This could describe a dual-mode optical waveguide, in
which inhomogeneities (e.g.~a Bragg grating) induce mixing between the
two types of transverse modes, as well as between forward-going and
backward-going modes.  It could also be realized using a
generalization of Mathur's quantum Hall model \cite{mathur1997}
to the quantum spin Hall gas \cite{Kane}.  In a quasi-one-dimensional
quantum spin Hall waveguide, shown schematically in
Fig.~\ref{fig:n2plots}(a), the coupled edge states can be described by
the Schr\"odinger equation
\begin{equation}
\begin{pmatrix}
-i\partial_z & -\eta & -m_1 & -a\\
-\eta^{*} & -i\partial_z & -a & -m_2\\
-m_1^{*} & -a^{*} & i\partial_z & -\eta^{*}\\
-a^{*} & -m_2^{*} & -\eta & i\partial_z
\end{pmatrix}\begin{pmatrix}
\psi_{\uparrow}^{+}\\
\psi_{\downarrow}^{+}\\
\psi_{\downarrow}^{-}\\
\psi_{\uparrow}^{-}
\end{pmatrix}
=\mathcal{E}\begin{pmatrix}
\psi_{\uparrow}^{+}\\
\psi_{\downarrow}^{+}\\
\psi_{\downarrow}^{-}\\
\psi_{\uparrow}^{-}
\end{pmatrix},
\label{spin_hall_model}
\end{equation}
where $\{\uparrow,\downarrow\}$ denote the spin polarizations of the
various edge states; $m_1$ and $m_2$ represent the scattering
amplitudes for spin-flip back-scattering along each edge; $\eta$
represents hopping between edges with spin flip; and $a$ represents
hopping between edges without spin-flip.  This Schr\"odinger equation
is invariant under the time-reversal operation $\psi_\uparrow^+
\leftrightarrow (\psi_\downarrow^-)^*$ and $\psi_\uparrow^-
\leftrightarrow (\psi_\downarrow^+)^*$, which conjugates the wave
amplitudes as well as reversing the spin and the direction of motion.
Re-arranging Eq.~(\ref{spin_hall_model}) yields the non-Hermitian
matrix $H$ given in Eq.~(\ref{H22a}) as the generator of the transfer
matrix.

Returning to Eq.~(\ref{H22a}), let us examine the possible eigenvalues
of $H$.  There are two special cases: first, if $m_1 = m_2 = 0$, then
the eigenvalues are
\begin{equation}
  \kappa=\pm \sqrt{(\mathcal{E}\pm|\eta|)^{2}-|a|^2},
\end{equation}
where the two $\pm$ signs are independent.  Secondly, if $\eta=0$ and
$m_1 = m_2 = m$, then the eigenvalues are
\begin{equation}
  \kappa=\pm\sqrt{E^{2}-|a\pm m|^2}.
\end{equation}
In both of these special cases, the eigenvalues are either real or
purely imaginary, and the transitions occur at $\kappa = 0$.  These
are the transitions between cases (i) and (ii) discussed in
Section~\ref{Eigenvalue properties}, and are essentially similar to
the behavior seen in the $N=1$ waveguide.  They correspond to passing
through band extrema at $k=0$.

When $\eta \ne 0$ and $m_1, m_2 \ne 0$, there can also occur
transitions between the cases (i) and (iii), or between (ii) and
(iii), discussed in Section~\ref{Eigenvalue properties}.  In other
words, the eigenvalues go from purely real or purely imaginary numbers
to a set of four distinct complex numbers.  We can demonstrate such
transitions by taking $a = 0$ and $m_1 = m_2 = m$, and varying $\eta$
and/or $\mathcal{E}$.  It can be shown that bifurcations occur when
\begin{equation}
  \mathcal{E} = \pm \Big|m \sin[\mathrm{arg}(\eta)]\Big|.
\end{equation}
They occur along the real-$\kappa$ axis if $|\eta| >
\big|m\cos[\mathrm{arg}(\eta)]\,\big|$, and along the
imaginary-$\kappa$ axis if $|\eta| <
\big|m\cos[\mathrm{arg}(\eta)]\,\big|$.  The former corresponds to
passing through a pair of band extrema located at $\pm \kappa$.
(Similar bifurcations occur if $a \ne 0$; the expressions for the
bifurcation points are simply more complicated.)  As in the $N=1$
case, we can observe the eigenvalue transitions using a waveguide
segment with varying parameters.  In Fig.~\ref{fig:n2plots}, we study
a waveguide segment in which $m_1 = m_2$ varies from zero (no
back-reflection) to a non-zero value.  Depending on the choice of
$\mathrm{arg}(\eta)$, $\kappa$ may remain real, or undergo a
real-to-complex bifurcation.  Unlike the $N=1$ case, however, this
bifurcation does not correspond to a breaking of an eigenstate
symmetry.  In the results obtained by backward integration of the
non-Hermitian Schr\"odinger-like equation, we indeed observe
exponential damping in the former case, and beating in the latter.
For $N > 2$, we expect to see the same relationships between the
reflection/transmission behavior of the waveguide and the underlying
non-Hermitian transitions of $H$, based on the eigenvalue properties
discussed in Section~\ref{Eigenvalue properties}.

\section{Conclusions}

We have shown that the generator of the transfer matrix, when regarded
as a non-Hermitian Hamiltonian, exhibits the features of
pseudo-Hermiticity
\cite{mostafa2002,mostafa2003,mostafa2004,mostafa2010} and anti-PT
symmetry \cite{PengPeng2016,Sukhorukov2015,LiGe2013,Wu2015}.  In the
literature on non-Hermitian systems, there has been a great deal of
interest in such symmetries as generalizations of the PT symmetry
concept~\cite{Bender,Bender2,Berry}.  In previous works, realizing
these symmetries has required the presence of phenomena such as
negative-index materials \cite{LiGe2013} or parametric amplification
\cite{Sukhorukov2015}.  In our case, they arise from the simple
physical requirements of flux conservation and time-reversal symmetry.
These Hamiltonians' non-Hermitian transitions, including the breaking
of anti-PT symmetry, manifest physically as the effects of crossing a
band extremum in a waveguide (e.g., the sharp decrease in transmission
when entering a bandgap).

We are grateful to M.~V.~Berry, J.~Gong, D.~Leykam, H.~Wang, and
Q.~Wang for helpful comments.  We acknowledge support from the
Singapore MOE Academic Research Fund Tier 2 Grant
No. MOE2015-T2-2-008, and the Singapore MOE Academic Research Fund
Tier 3 Grant MOE2011-T3-1-005.

\end{document}